\documentclass[preprint]{aastex}

\usepackage{url}

\shorttitle{iPTF Realtime Image Subtraction Pipeline}
\shortauthors{Cao et al.}

\begin{document}

\title{Intermediate Palomar Transient Factory: Realtime Image
  Subtraction Pipeline}

\author{Yi Cao\altaffilmark{1}, Peter~E.~Nugent\altaffilmark{2,3} \&
  Mansi~M.~Kasliwal\altaffilmark{1}} \altaffiltext{1}{Astronomy
  Department, California Institute of Technology, Pasadena, CA 91125,
  USA} \altaffiltext{2}{Department of Astronomy, University of
  California, Berkeley, CA 94720-3411, USA} \altaffiltext{3}{Lawrence
  Berkeley National Laboratory, 1 Cyclotron Road, MS 50B-4206,
  Berkeley, CA 94720, USA}

\begin{abstract}
A fast-turnaround pipeline for realtime data reduction plays an
essential role in discovering and permitting follow-up observations to
young supernovae and fast-evolving transients in modern time-domain
surveys. In this paper, we present the realtime image subtraction
pipeline in the intermediate Palomar Transient Factory. By using
high-performance computing, efficient database, and machine learning
algorithms, this pipeline manages to reliably deliver transient
candidates within ten minutes of images being taken. Our experience in
using high performance computing resources to process big data in
astronomy serves as a trailblazer to dealing with data from
large-scale time-domain facilities in near future.
\end{abstract}


\section{Introduction}
\label{sec:intro}
Transients on short timescales ($\lessapprox1\,\textrm{day}$) are
becoming particularly interesting for at least two reasons: first,
radiation from a supernova (SN) in the first few days of explosion
contains rich information about its initial condition, including the
progenitor star and its environment \citep[e.g., ][]{k10, ns10, rw11}.
Hence observations of young SNe open a new phase space to constrain
the poorly understood physical properties of the exploding stars in
their last evolutionary stages
\citep[e.g.,][]{sbp+08,agy+11,nsc+11,cka+13,ckh+15,gtr+16}.  Second,
some relativistic transient phenomena have been found to decay on
sub-day timescales \citep[e.g.,][]{ckh+13,cup+15}. These extreme
events are probably related to deaths of the most massive stars.

In recent years, many time-domain projects are motivated to perform
fast-cadence transient surveys at different scales, e.g., SDSS
\citep{sbb+08}, Pan-STARRS \citep{rsf+14}, the intermediate Palomar
Transient Factory (iPTF) and ASAS-SN.  Since 2013, as the successor of
PTF \citep{lkd+09}, iPTF have been performing several fast-cadence
experiments to systematically search for and characterize young SNe
and fast-evolving transients. Since the survey itself only provides
single-band monitoring, detailed multi-wavelength follow-up
observations are warranted to collect additional information in order
to examine the nature of discovered transients, such as
ultraviolet-optical-infrared light curves and colors, spectroscopic
classification, X-ray and radio observations for non-thermal emission,
etc.. As these transients evolve rapidly, to allow time to undertake
follow-up observations, we developed an automated image subtraction
pipeline that reliably identifies interesting transients soon after
the survey images are taken.

This paper is organized as follows: a brief summary of the
fast-cadence surveys in iPTF is in \S\ref{sec:iptf}.  The pipeline and
its performance is described in
\S\ref{sec:pipeline}. \S\ref{sec:summary} summarizes the paper.

\section{A Brief Summary Of iPTF}
\label{sec:iptf}
iPTF uses the 48\,inch Schmidt telescope (P48) at the Palomar
Observatory which is equipped with the CFH12K camera \citep{slc+00} as
the discovery machine to monitor the sky. The CFH12K camera has a
mosaic of eleven working CCDs, each of which has $2048\times4096$
pixels with a pixel size
$1.01^{\prime\prime}\,\textrm{pixel}^{-1}$. The total field of view is
therefore $7.26\,\textrm{deg}^2$. The camera has two filter options:
Mould \textit{R} ($\lambda_{\textrm{center}}=6581\,\textrm{\AA}$,
$\textrm{FWHM}=1251\,\textrm{\AA}$) and \textit{g}
($\lambda_{\textrm{center}}=4754\,\AA$,
$\textrm{FWHM}=1537\,\textrm{\AA}$).  The exposure time is currently
chosen to be 60 seconds so a single snapshot reaches a depth of
$R\simeq20.5$\,mag or $g\simeq21\,$mag under the median seeing of
$2^{\prime\prime}.0$ at the Palomar Observatory. These limiting
magnitudes are set to match our spectroscopic follow-up
capability. The mean overhead time of readout and telescope slew for
each exposure is about 40 seconds.

In the spring and fall observing seasons, iPTF performs fast-cadence
experiments to search for young SNe and fast transients. Each field is
visited multiple times every night in one of the two filters. Once a
frame of 11 individual CCD images (file size: $176\,$MB) is read out,
it is immediately transferred to the National Energy Research
Scientific Computing Center (NERSC) for further processing where our
realtime image subtraction pipeline is running.  iPTF usually takes
$\simeq300$ (in summer) to $>400$ (in winter) exposures every night.
So the nightly data volume is 50--70\,GB.

\section{Realtime Image Subtraction Pipeline}
\label{sec:pipeline}

\begin{figure}[th]
\centering \includegraphics[width=\textwidth]{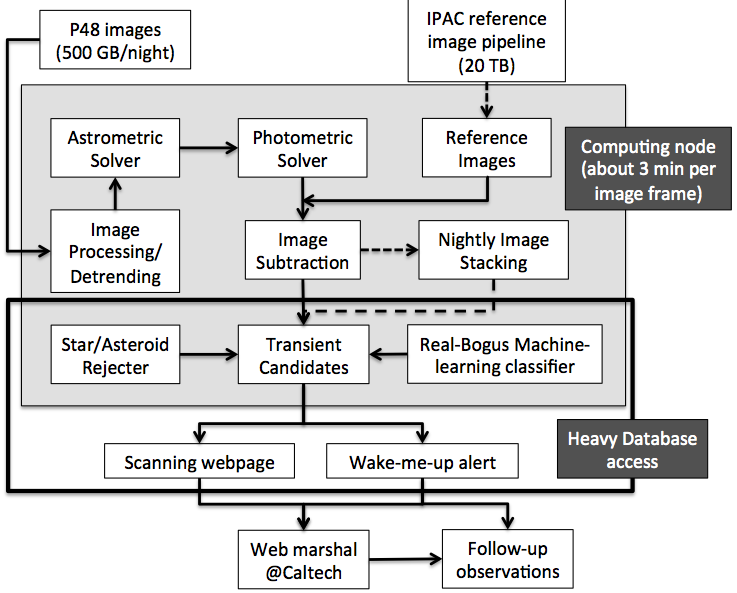}
\caption[Overview of the NERSC image subtraction pipeline] {Overview
  of the NERSC image subtraction pipeline. The gray region highlights
  the realtime pipeline running on the NERSC supercomputer. The black
  box highlights intercommunication to the database. Solid arrows
  represent realtime operations and dashed arrows represent post-night
  operations in the pipeline. This paper focuses on the
  gray-highlighted region. }
\label{fig:pipeline}
\end{figure}

A flow chart of the realtime image subtraction pipeline is shown in
Figure \ref{fig:pipeline}.

On top of the pipeline is a robotic scheduler. It identifies incoming
image frames on the spinning disk, creates a database record for each
image frame, and submits one processing job for each frame to the
supercomputer queue. It is also responsible for post-processing
procedures at the end of a night, including backup of raw and
processed data, starting the post-night image stacking procedure, and
sending out nightly processing reports.

Each processing job is placed on one computing node with multiple
CPUs. It generates a child process for each individual CCD image which
utilizes one CPU core and about 4GB of memory. With the exception of a
serial run for the initial astrometric solution, all CCD images from a
given frame are processed in parallel on a computing node.  In the
following, we describe the major steps in the image subtraction
pipeline:

\begin{itemize}
\item \textbf{Image preprocessing --} We perform routine procedures of
  overscan subtraction, bias subtraction and flat-field correction to
  the raw image. Since bias and flat-field features vary on long
  timescales, a separate daytime program updates the bias and
  flat-field images every ten days. For each CCD, this program
  median-stacks the most recent 100 bias images to make super biases,
  and the most recent 100 science images in $g$ or $R$ filters to
  generate super flats in the corresponding filters. Bad-pixel masks
  are also updated at the same time.

\item \textbf{Astrometric and photometric solvers --} Given that each
  CCD covers $0.57^\circ\times1.15^\circ$ sky area,
  \texttt{astrometry.net}\footnote{\texttt{astrometry.net} is
    available at \url{http://astrometry.net}.} is well suited to
  derive an initial astrometric solution. The success rate of
  \texttt{astrometry.net} is $>99\%$ and the resulting root mean
  square of the astrometric error is less than half a pixel. However,
  the current version of \texttt{astrometry.net} does not support
  parallelization. Running a separate instance of
  \texttt{astrometry.net} for each CCD child process would require an
  extremely large mount of memory and could also be limited by the
  memory access speed. As such we are forced to synchronize all child
  processes after image preprocessing and run \texttt{astrometry.net}
  on a single CPU core.

After solving for the astrometric solution, each child process uses
\texttt{SExtractor}\footnote{\texttt{SExtractor} is available at
  \url{http://www.astromatic.net/software/sextractor}.} to extract a
catalog of sources with signal-to-noise ratios (SNRs) $>20$. This
catalog is spatially matched to the PTF-iPTF catalog \citep{ols+12} to
calibrate the photometry. To make this step fast, only a median
zero-point across each CCD image is calculated, although some CCD
images have zero-point variation at different pixels \citep{oll+12}.
The precision of the photometric solution at this stage is about
$0.1\,$mag at 20th magnitude.

\item \textbf{Image registration --} In preparation for image
  subtraction, we need to register both the science image and its
  corresponding reference image into the same pixel coordinate
  system. Here the reference image is generated by stacking several
  best-quality images of the same field. This task is performed by
  IPAC which is in charge of PTF/iPTF data archiving. Hence, the noise
  level in the reference image is usually negligible compared to that
  in the science image. So we remap the reference image into the pixel
  coordinate system of the science image.  First, we generate a
  catalog of sources with SNR$>5$ on the science image and a catalog
  of sources on the reference image. In order to balance the numbers
  of sources between the science and reference catalogs for more
  precise alignment, we truncate the reference catalog at the
  5$\sigma$ limiting magnitude of the science image. Then the science
  catalog is matched spatially to the reference catalog with
  \texttt{Scamp}\footnote{\texttt{Scamp} is available at
    \url{http://www.astromatic.net/software/scamp}.}.  Since the
  reference image always has an astrometric precision
  $\lessapprox0.1^{\prime\prime}$, and since the \texttt{Scamp}
  matching usually yields a rms $\lessapprox0.1^{\prime\prime}$, the
  final astrometric solution on the new image is
  $\lessapprox0.2^{\prime\prime}$.  After solving the astrometry of
  the science image with respect to the reference image, we remap the
  reference image into the pixel coordinate system of the science
  image with \texttt{SWarp}\footnote{\texttt{SWarp} is available at
    \url{http://www.astromatic.net/software/swarp}.}.

\item \textbf{Image subtraction --} In this step, we subtract the
  remapped reference image from the science image with
  \texttt{HOTPANTS}\footnote{\texttt{HOTPANTS} is available at
    \url{http://www.astro.washington.edu/users/becker/v2.0/hotpants.html}}.
  \texttt{HOTPANTS} is an implementation of the image subtraction
  algorithm by \citet{al98}.  In order to minimize the error on the
  resulting subtraction image, the PSF on the remapped reference image
  is converted into the PSF on the new image by a PSF convolution
  kernel. It can be proved in the limiting case of a noise-free
  reference image that this method is the most optimal solution for
  background-limited images \citep{zog16}. The convolution kernel is
  approximated by linear combinations of three Gaussian
  functions---one of which has a width narrower than the seeing, a
  second one has a width similar to the seeing, and the third one has
  a width wider than the seeing. The linear coefficients are
  calculated by using a list of high-SNR stars which are selected from
  the \texttt{SExtractor} catalogs and which are evenly distributed
  across both images. Low spatial-frequency variation of the PSF
  across the image is also modeled by low-order polynomials in
  \texttt{HOTPANTS}.  Table \ref{tab:hotpants} lists our
  \texttt{HOTPANTS} input parameters.  After image subtraction, the
  flux level of the subtracted image is normalized to that of the
  reference image. The subtracted image usually reaches a 5$\sigma$
  limiting magnitude of $R\sim20.5$\,mag or $g\sim21$\,mag.  After
  image subtraction, we run \texttt{SExtractor} to look for all
  sources with SNR$>5$ as transient candidates.  Typically,
  \texttt{SExtractor} finds a couple of hundred candidates on each
  subtracted image of an extragalactic field.

\begin{deluxetable}{ccc}
  \tablecaption{\texttt{HOTPANS} parameters}\label{tab:hotpants}
  \tablehead{\colhead{Parameter} & \colhead{Value} & \colhead{Note}}
  \startdata r & $2.5\times\textrm{seeing}$ & convolution kernel half
  width (pixel) \\ rss & $6\times\textrm{seeing}$ & half width of
  substamp to extract around stars (pixel) \\ tu & recorded in the
  reference header & upper valid data count of the reference
  image\\ tl &
  median(reference)$-10\times\sigma_{\textrm{reference}}$\tablenotemark{1}
  & lower valid data count of the reference image\\ iu & 45000 & upper
  valid data count of the science image\\ il &
  median(science)$-10\times\sigma_{\textrm{science}}$\tablenotemark{1}
  & lower valid data count of the science image\\ \enddata
  \tablenotetext{1}{$\sigma$ denotes the median absolute deviation.}
\end{deluxetable}

\item \textbf{Real-bogus classification --} Nonlinearity of the
  detectors, astrometric misalignment, imperfect convolution kernels,
  Poisson noise of bright objects, cosmic rays, and many other factors
  may produce artifacts on the subtracted image. To remove these
  image-based artifacts, we developed a machine-learning real-bogus
  classifier which uses the random forest algorithm. For each
  candidate, we defined 42 features using pixels from new, reference
  and subtracted images. These features reflect information about the
  candidate shape, the quality of the subtraction, nearby sources and
  so on. The ``real'' training set consists of variable stars,
  spectroscopically classified supernovae, and asteroids, while the
  ``bogus'' set are a large number of randomly selected candidates in
  our database and confirmed by visual inspection, which represents
  the whole false positive population. This classifier achieves a
  $1\%$ false positive rate at a cost of $5\%$ false negative rate
  \citep{brp+13,rbw15}. In practice, only the top few percent of the
  candidates in each subtracted image may pass this real-bogus
  classifier.

\item \textbf{Matching external catalogs --} After removing most
  artifacts in the transient candidate list, the main contaminating
  sources are real celestial objects, including asteroids, variable
  stars and active galaxy nuclei (AGNe). As such we further match the
  transient candidates to the minor planet catalog to remove
  asteroids. We also use the SDSS catalogs to identify variable stars
  and AGNe. Outside the SDSS footprint, we built our own star catalog
  by applying the following criteria to the reference
  \texttt{SExtractor} catalog:
\begin{itemize}
\item MAG\_BEST$<$20;
\item
  $\textrm{MU\_MAX}-\textrm{MAG\_BEST}<0.2+\textrm{Median}(\textrm{MU\_MAX}-\textrm{MAG\_BEST})$;
\item
  $\textrm{FWHM\_IMAGE}<2\times\textrm{Median}(\textrm{FWHM\_IMAGE})$;
\item $\textrm{ELONGATION} < 1.5$\,
\end{itemize}
where $\textrm{Median}(*)$ is the median function for the whole
\texttt{SExtractor} catalog. This star-galaxy classifier is able to
identify $50\%$ of the field stars and meanwhile misclassifies $5\%$
of galaxies as stars. Very recently, Miller et al. (in prep.)
developed a new random-forest star-galaxy classifier for the PTF/iPTF
data. The new classifier uses the whole set of \texttt{SExtractor}
outputs and is trained by spectroscopically confirmed stars in
SDSS. It achieves a higher true positive rate and a much lower false
positive rate.

Matching transient candidates to external catalogs also allows us to
assess the quality of a subtracted image by using the following two
metrics: the ratio between the total numbers of transient candidates
and stars, denoted as $\xi$, and the fraction of transient candidates
that pass the real-bogus test (the real-bogus score threshold
corresponds to a $1\%$ false positive rate and $5\%$ false negative
rate), denoted as $\eta$. If we make two reasonable assumptions that
the fraction of variable stars is almost constant (denoted as $\phi$)
at different parts of the sky and that half of the variable stars are
brighter in the science image than in the reference image, then $\xi$
should roughly equal to $\phi/2$. If we further assume that real
transients are much rarer than variable stars, then $\eta$ should be
close to zero. Subtracted images with large $\xi$ and $\eta$ values
usually result from substantial misalignment between the science and
reference images.

In addition to point source catalogs, we also spatially associate
transient candidates to nearby galaxies in the CLU catalog (Cook et
al. in prep.), as the absolute magnitude and environment of a
transient also provide valuable information about the nature of the
transient.

\item \textbf{Post-night image stacking --} Thanks to the multiple
  visits to the same field each night, after each night, we stack the
  subtracted images of each field with \texttt{Swarp} to deepen our
  detection limits by about half a magnitude. These stacked images are
  used to search for transients slightly below the single-image
  detection limit, to provide better SNRs for detected transients in
  single images, and to provide deeper detection limits for
  non-detection.

\end{itemize}

In all these steps, the pipeline intensively communicates with our
transient database which (1) records metadata of incoming images, (2)
searches for reference images, (3) stores metadata of subtracted
images and transient candidates, and (4) matches transient candidates
to external catalogs. In order to facilitate the database, we take the
advantage of the NERSC high-performance database service and design a
database schema to minimize JOIN operations between big tables.

\begin{figure}
\centering
\includegraphics[width=0.95\textwidth]{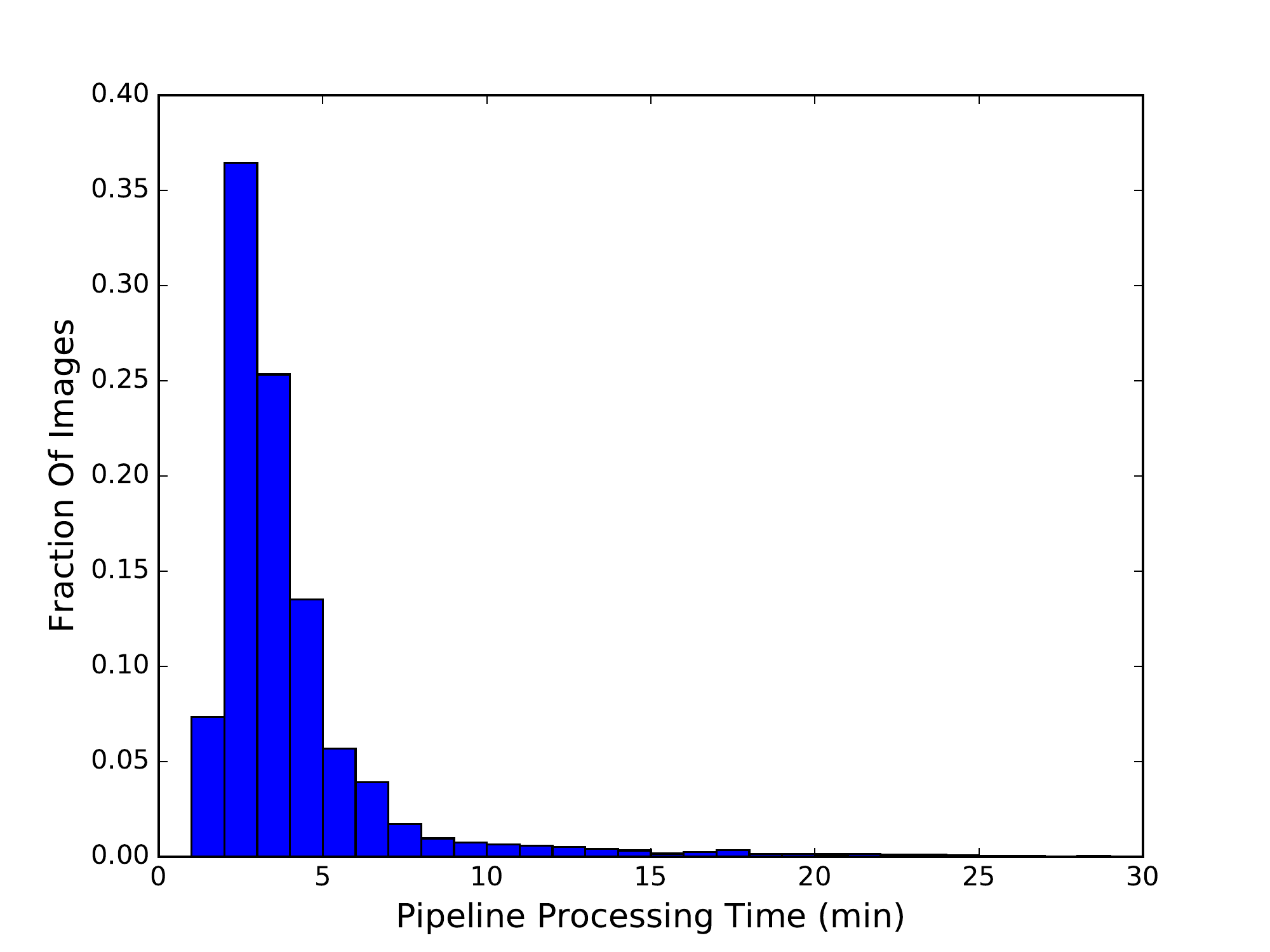}
\caption[Histogram of NERSC pipeline processing time]{Histograms of
  NERSC pipeline processing time.}
\label{fig:processTime}
\end{figure}

Table \ref{tab:elapsedTime} shows the 10th percentile, median, and
90th percentile of elapsed times of the individual steps and the
pipeline as a whole. The median total elapsed time of a processing job
is $3.1$ minutes and its distribution is illustrated in the histogram
of Figure \ref{fig:processTime}.

\begin{figure}
\centering \includegraphics[width=0.95\textwidth]{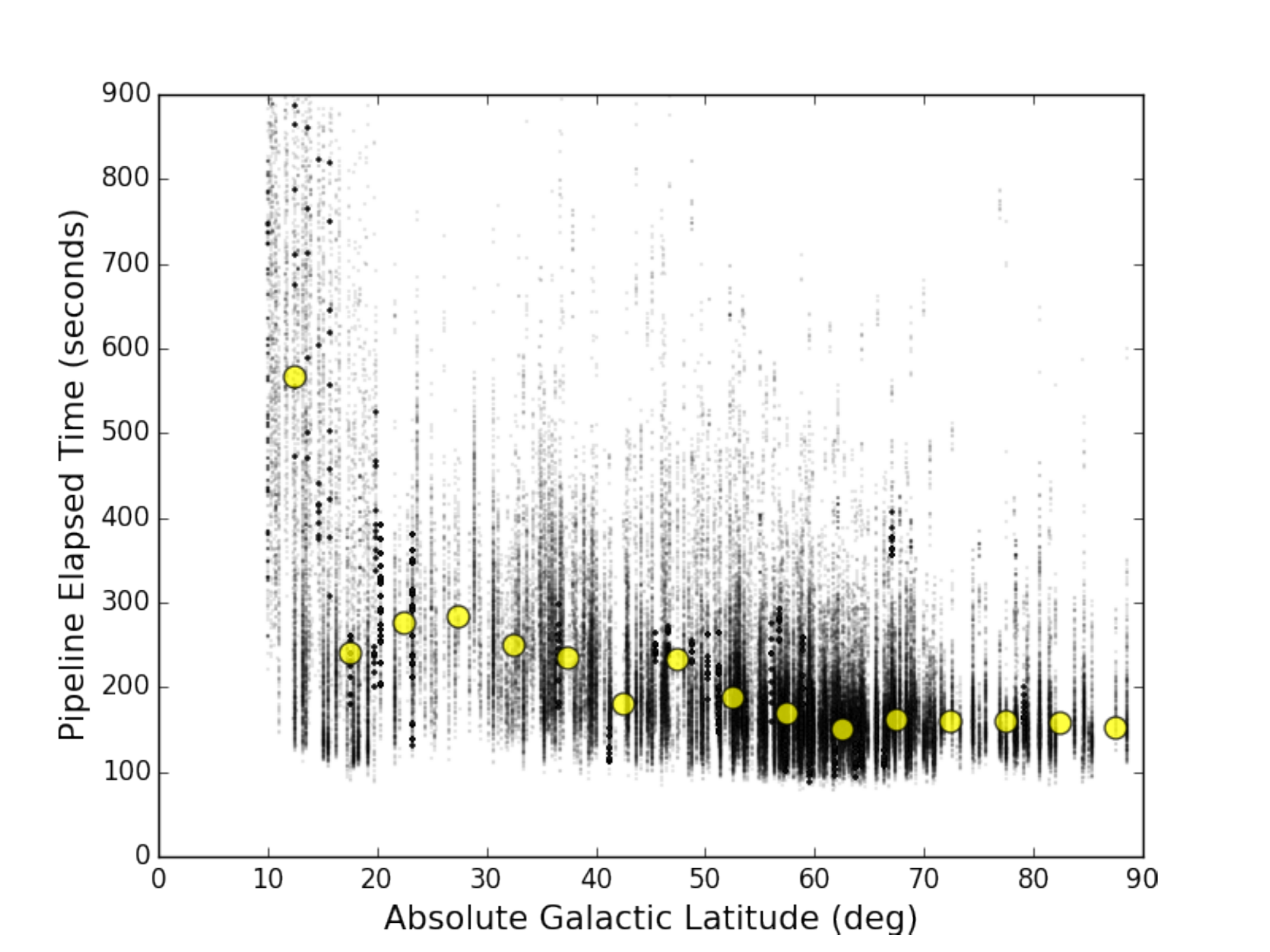}
\caption{Elapsed time of the pipeline as a function of galactic
  latitude. The gray dots are real measurements and the yellow circles
  denote the median values in bins of 5 degrees in the absolute
  galactic latitudes. Note that, since the primary goal of our
  pipeline is to search for extragalactic transients, it does not
  process any image within 10 degrees of the Galactic plane.
\label{fig:galacticLatitude}}
\end{figure}

We also note that the total elapsed time increases dramatically at low
galactic latitudes (Figure \ref{fig:galacticLatitude}). This is
because the stellar density is high at low galactic latitudes. Any
operation involving real celestial sources---such as detecting all
sources on an image and matching sources to external catalogs---takes
longer at higher stellar densities.

\begin{deluxetable}{cccc}
	\centering \tablewidth{0pt} \tablecolumns{4}
        \tablecaption{Elapsed Times of Pipeline
          Steps\label{tab:elapsedTime}} \tablehead{\colhead{Pipeline
            Steps} & \multicolumn{3}{c}{Elapsed Time (seconds)}
          \\ \colhead{} & \colhead{10th percentile} & \colhead{median}
          & \colhead{90th percentile}} \startdata Preprocessing &
        $2.8$ & $3.5$ & $7.3$ \\ Astrometry and photometry solver &
        $11.9$ & $16.3$ & $42.2$ \\ Image registration & $18.2$ &
        $23.2$ & $44.9$ \\ Image subtraction & $24.1$ & $29.1$ &
        $44.2$ \\ Real-bogus classification & $23.8$ & $40.7$ & $84.3$
        \\ Matching external catalogs & $28.9$ & $61.3$ & $168.9$
        \\ \hline Pipeline total & $125.6$ & $190.8$ & $379.4$
        \\ \enddata
\end{deluxetable}

\begin{figure}[th]
\centering
\includegraphics[width=0.9\textwidth]{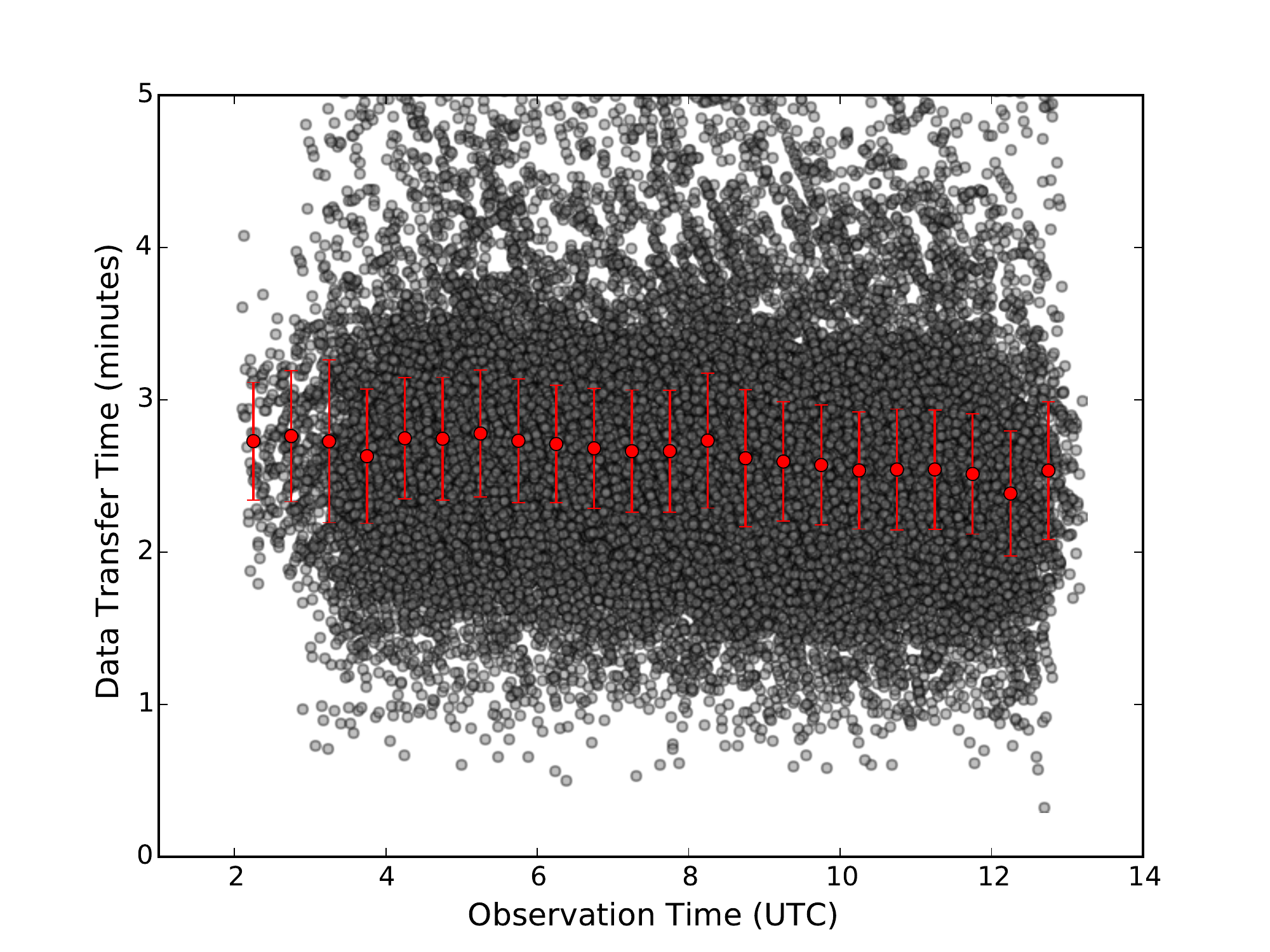}
\caption{The transfer time of individual image frames as a function of
  UTC hours. The red circles denote the median data transfer time in
  half-an-hour bins and the red error bars are the median absolute
  deviation in the corresponding bins.
\label{fig:transferTime}}
\end{figure}

Next, we consider the overhead time of processing images after images
have been taken.  The data transfer from the Palomar Observatory to
NERSC uses a microwave link between the Palomar Observatory and the
San Diego Supercomputer Center and ESnet\footnote{ESnet is the
  Department of Energy's dedicated science network. More information
  can be found at \url{https://www.es.net/}.} between the San Diego
Supercomputer to NERSC through Caltech. It usually takes 2-3 minutes
(Figure \ref{fig:transferTime}) for the images from the Palomar
Observatory to NERSC. There is a weak trend that the data transfer is
slightly faster in the second half of the night than in the first
half, probably due to decreasing use by people during the early
morning hours.

Upon receipt of each image frame, it usually takes up to half a minute
for the scheduler to locate this incoming frame. Then the schedule
spends seconds on performing a validation check on the image header,
including (1) whether the image frame is a science frame, (2) whether
it is taken in either the \textit{R} or \textit{g} filter, and (3)
whether the galactic latitude of the field is above 10 degrees from
the Galactic plane. After the image frame passes this validation
check, the scheduler generates and submits a job to the supercomputer
queue to process it.

The common supercomputer queues are usually filled with jobs from all
users and therefore have a queue waiting time ranging from a few
minutes to a few hours. To reduce the queue waiting time, NERSC has
set up a special realtime queue with eight dedicated computing nodes.
Our jobs in this queue are usually executed within half a minute of
submission. Occasionally when there are more than eight jobs in the
queue, the first eight jobs get executed and other jobs are put on
hold for the next available node.

\begin{figure}[th]
\centering \includegraphics[width=0.9\textwidth]{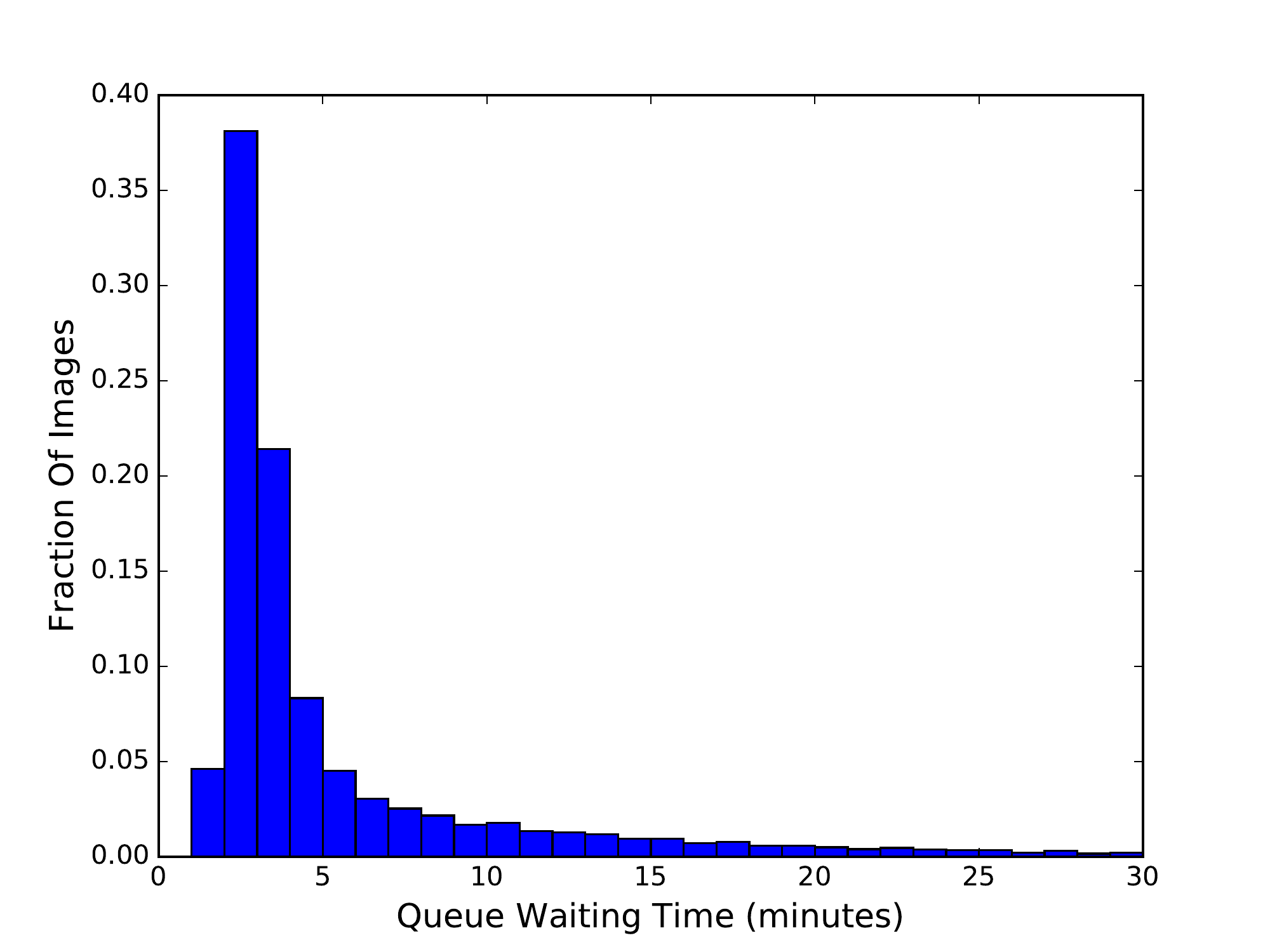}
\caption{The distribution of wait time between receipt of an image
  frame at NERSC and the execution of its processing job.
\label{fig:waitTime}}
\end{figure}

The total wait time between receipt of an image frame and execution of
its corresponding process job is shown in Figure \ref{fig:waitTime}.

\begin{figure}
\includegraphics[width=0.95\textwidth]{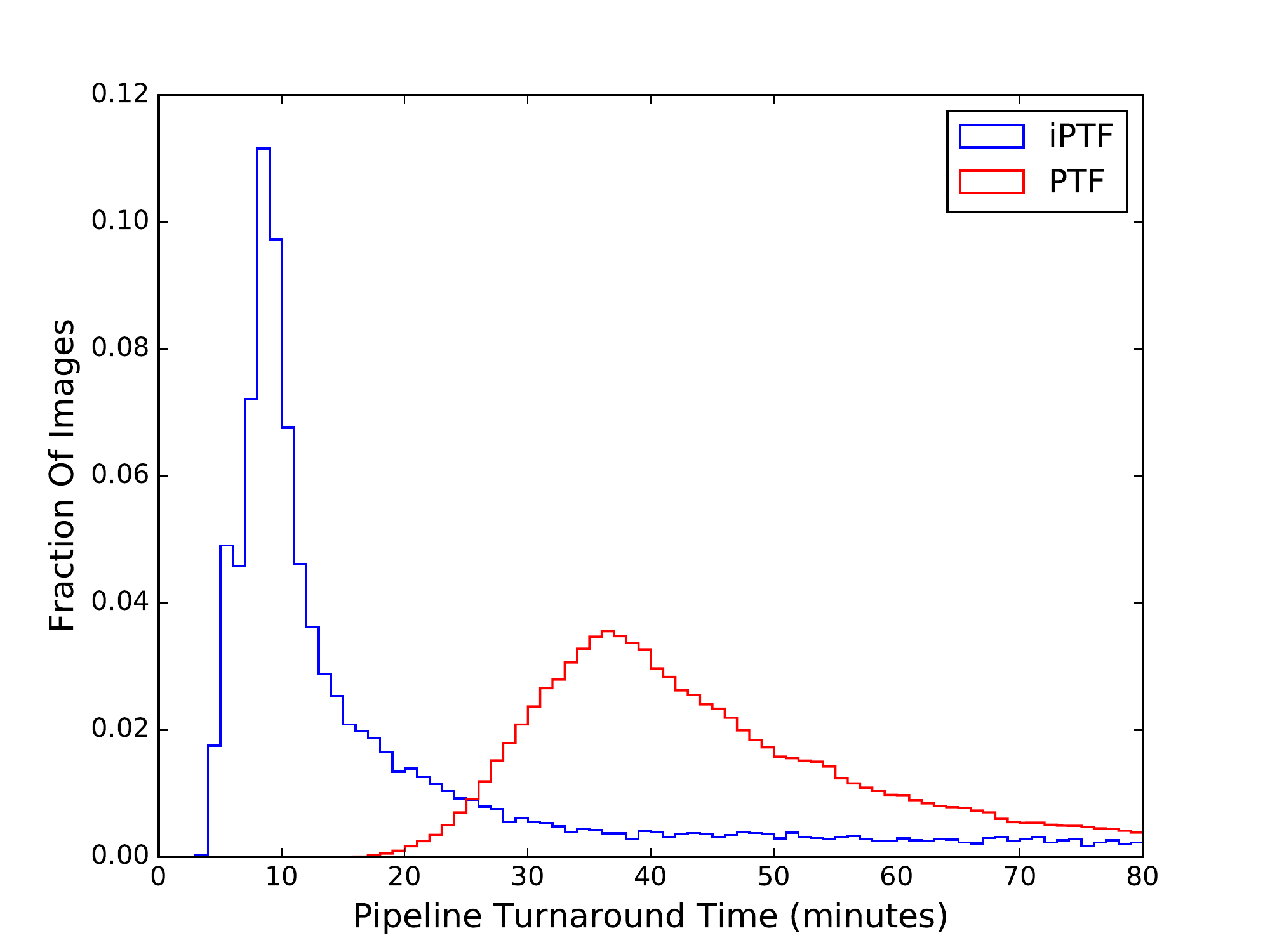}
\caption[Histogram of turnaround time from completion of an exposure
  to output candidates from the pipeline] {Histogram of turnaround
  time from completion of an exposure to output candidates from the
  realtime pipeline in PTF (red) and iPTF (blue).}
\label{fig:turnaroundTime}
\end{figure}

With the data transfer and wait times included, the total turnaround
time from completion of an exposure to candidates with real-bogus
scores in the database is shown in Figure \ref{fig:turnaroundTime}.
As can be seen in the figure, the whole pipeline manages to deliver
transient candidates within ten minutes of most images being taken.

Compared to the PTF pipeline, our iPTF pipeline reduced the turnaround
time by a factor of $3\sim4$. This speedup results from improvements
in both hardware and software. For the hardware, we are provided
sufficient computing nodes and lustre filesystems with fast I/O. On
the software side, we properly designed an interface to the database
which records the timing of all steps in the pipeline. These records
allow us to locate the bottlenecks of the pipeline and concentrate our
man power on them (e.g., indexing the database tables to speed up
certain queries).

This fast turnaround pipeline has enabled the iPTF collaboration to
discover and study over a hundred young SNe and several fast-evolving
transients. The related publications thus far are listed in Table
\ref{tab:publications}.

\begin{deluxetable}{cll}
	\centering \tablecolumns{3} \tablewidth{0pt}
        \tablecaption{iPTF publications on young SNe and fast-evolving
          transients\label{tab:publications}} \tablehead{
          \colhead{Objects} & \colhead{Publications} &
          \colhead{Titles} } \startdata 57 Type II Supernovae &
        \citet{rgd+16} & Type II Supernova Energetics and Comparison
        of \\ & & Light Curves to Shock-cooling Models \\ 84 Type II
        Supernovae & \citet{kyg+16} & Flash Spectroscopy: Emission
        Lines from the Ionized \\ & & Circumstellar Material around
        $<10$-day-old Type II \\ & & Supernovae \\ iPTF14atg &
        \citet{ckh+15} & A strong ultraviolet pulse from a newborn
        type Ia \\ & & supernova \\ iPTF13ebh & \citet{hbc+15} &
        Strong near-infrared carbon in the Type Ia supernova\\ & &
        iPTF13ebh \\ iPTF14yb & \citet{cup+15} & iPTF14yb: The First
        Discovery of a Gamma-Ray Burst \\ & & Afterglow Independent of
        a High-energy Trigger \\ iPTF13beo & \citet{ggo+14} &
        iPTF13beo: the double-peaked light curve of a Type Ibn \\ & &
        supernova discovered shortly after explosion \\ iPTF13ast &
        \citet{gao+14} & A Wolf-Rayet-like progenitor of SN 2013cu
        from\\ & & spectral observations of a stellar wind
        \\ iPTF13bvn & \citet{fst+14} & The rise and fall of the Type
        Ib supernova \\ & & iPTF13bvn. Not a massive Wolf-Rayet star
        \\ iPTF14jj & \citet{gja+14} & The Rise of SN 2014J in the
        Nearby Galaxy M82 \\ iPTF13bvn & \citet{cka+13} & Discovery,
        Progenitor and Early Evolution of a \\ & & Stripped Envelope
        Supernova iPTF13bvn \\ \enddata
\end{deluxetable}

\section{Summary and Discussions}
\label{sec:summary}
In this paper, we present the iPTF realtime image subtraction pipeline
which, by properly combining high-performance computing resources at
NERSC, existing astronomical software, databases, and machine learning
algorithm, manages to reliably deliver transient candidates within ten
minutes of images being taken at the Palomar Observatory. This
pipeline makes identification of interesting fast-evolving transients
as fast as possible and saves time for us to undertake follow-up
observations. Our experience demonstrates that, in a modern transient
survey, data reduction software which follows the survey observations
and enables follow-up observations is equally important as the survey
and follow-up observations themselves.

A few technical issues in the current pipeline could be improved in
future pipelines.  First, the pipeline utilizes existing astronomy
software which reads input from and passes results onto spinning
disks. Therefore, when a large volume of data are being read or
written simultaneously, or when the I/O speed of the spinning disks is
limited for other reasons, the processing can be substantially slowed
down.  Furthermore, the I/O speed of the spinning disks is
intrinsically much slower than that of memory. Hence, the future
pipeline should perform most operations in memory or using
BurstBuffer\footnote{More information about BurstBuffer can be found
  at
  \url{http://www.nersc.gov/users/computational-systems/cori/burst-buffer/}.}
and limit the communication to the spinning disks when necessary.

Second, the simple Gaussian approximation to the convolution kernel in
\texttt{HOTPANTS} is not perfect, because it fails to capture
non-Gaussian features of the PSF, especially when the observing
conditions are not photometric. In these cases, the image subtraction
may produce many subtraction residues of the subtracted images.  In
order to enhance the quality of image subtraction, several
improvements to the \citeauthor{al98} algorithm have been proposed and
tested (\citealt{zog16}; Masci et al. in prep.).

Third, Python is used as a high-level wrapper in the pipeline and also
performs simple algebraic calculations.  The initialization of Python
and execution is generally much slower than a machine-language
executable.  Loading Python modules also highly relies on the I/O
speed to the spinning disks. Thus, moving this to the BurstBuffer
should also make significant speed-ups.

Moving forward, iPTF will cease operation in February 2017 and make
way for the Zwicky Transient Facility (ZTF; \citealt{sdb+14,ztf}),
which will mount a new camera on the P48 telescope. The new camera
will have a field of view of 47 square degrees with faster readout.
The optimal survey speed in ZTF will be more than ten times faster
than that in iPTF.  ZTF will be able to survey the entire visible sky
down a depth similar to iPTF every eight hours. Therefore, ZTF is
poised to chart the phase space on sub-day timescales. Accordingly, we
are organizing a systematic global follow-up network, with the GROWTH
(Global Relay of Observatories Watching Transients Happen) program,
that is focused on fast transients, young supernovae and asteroids
within the first 24 hours of discovery.

Quick discovery pipelines that bridge discovery and rapid follow-up
observations will continue to be critical in the ZTF success. The data
rate in ZTF will be ten times higher than in iPTF, which is a certain
challenge to the ZTF pipeline. By coupling the astronomical pipelines
with advanced high-performance computing resources, we are well on our
way to delivering this capability for ZTF.

\acknowledgements We thank J. Sollerman for useful suggestions to
improve the manuscript.  YC and PEN acknowledge support from the DOE
under grant DE-AC02-05CH11231, Analytical Modeling for Extreme-Scale
Computing Environments.  YC and MMK also acknowledge support from the
National Science Foundation PIRE program grant 1545949.  The
intermediate Palomar Transient Factory project is a scientific
collaboration (PI: S. R. Kulkarni) among the California Institute of
Technology, Los Alamos National Laboratory, the University of
Wisconsin, Milwaukee, the Oskar Klein Center, the Weizmann Institute
of Science, the TANGO Program of the University System of Taiwan, and
the Kavli Institute for the Physics and Mathematics of the Universe.

\end{document}